\documentclass[preprint,pra,showpacs,nofootinbib]{revtex4}
\def\be{\begin{equation}}
\def\ee{\end{equation}}
\def\ben{\begin{eqnarray}}
\def\een{\end{eqnarray}}

\begin{document}

\begin{flushright}
\vspace*{1cm}
\end{flushright}

\title{\bf Vacuum fluctuations and Brownian motion of a charged test particle
near a reflecting boundary}

\author{Hongwei Yu}
\email{hwyu@cosmos.phy.tufts.edu}
\affiliation{ CCAST(World Lab.),
P. O. Box 8730, Beijing, 100080, P. R. China and Department of
Physics and Institute of  Physics,\\ Hunan Normal University,
Changsha, Hunan 410081, China\footnote{Mailing address}}

\author{L. H. Ford }
\email{ford@cosmos.phy.tufts.edu}
\affiliation{Institute of Cosmology, Department of Physics and Astronomy\\
    Tufts University, Medford, MA 02155, USA}

\begin{abstract}
We study the  Brownian motion of a charged test particle coupled
to electromagnetic vacuum fluctuations near a perfectly reflecting plane
boundary. The presence of the boundary modifies the quantum  fluctuations
of the electric field, which in turn modifies the motion of the test particle.
We calculate the resulting mean squared fluctuations in the velocity and
position of the test particle. In the case of directions transverse
to the boundary, the results are negative. This can be interpreted
as reducing the quantum uncertainty which would otherwise be present.
\end{abstract}

\maketitle
\baselineskip=13pt

\section{Introduction }

In quantum electrodynamics, the effects of electromagnetic vacuum fluctuations
upon an electron in empty space are usually regarded as unobservable. The
divergent parts of the electron self-energy are absorbed by mass and
wavefunction renormalizations, and the finite self-energy function can be taken
to vanish for real (as opposed to virtual) electrons. However, {\it changes}
in the vacuum fluctuations can produce observable effects. The Lamb shift
and the Casimir effect are two examples of this.

In the present paper, we wish to discuss a very simple situation, the Brownian
motion of a charged particle coupled to the quantized electromagnetic
field. Just as a classical stochastic field will cause random motion of a test
particle, one might also expect Brownian motion to be caused by quantum
fluctuations. It is unclear whether this motion can be observable in the
Minkowski vacuum state, although Gour and Sriramkumar~\cite{GS99} argue
that it might be. Jaekel and Reynaud~\cite{JR} have also discussed this issue
in the context of mirrors coupled to vacuum fluctuations.
 Here we will be concerned with shifts due to the quantum
state of the field being other than the Minkowski vacuum. One
simple way to cause a nontrivial shift in the vacuum fluctuations is to
introduce a reflecting boundary. In this paper, we will discuss the case
of a single, perfectly reflecting plate, and calculate the effects of the
modified electromagnetic vacuum fluctuations upon the motion of a charged
test particle. The analogous calculations for the case of an uncharged,
polarizable test particle were reported in Ref.~\cite{WKF}. The present problem
bears some analogy to the problem of lightcone fluctuations, where
photons undergo Brownian motion due to modified quantum fluctuations of the
quantized gravitational field~\cite{F95,YF99,YF00}.

\section{The Langevin equation and its Solutions}

We treat the particle as a point particle of mass $m$ and electric
charge $e$. In the limit of low velocities, the velocity ${\mathbf v}$
satisfies the non-relativistic equation of motion with only an electric
force term:
\begin{equation}
\frac{d {\mathbf v}}{d t} = \frac{e}{m}\, {\mathbf{E}}({\mathbf{x}},t)\,.
 \label{eq:Lang}
\end{equation}
We will restrict our attention to the case where the particle does not move
significantly, so we can assume the position $\mathbf{x}$ to be constant.
We also assume that dissipation can be ignored.
If the particle starts at rest at time $t=0$, then at time $t$ the velocity
is
\begin{equation}
 {\mathbf{v}}={e\over m}\int_0^t\;{\mathbf{E}}({\mathbf{x}},t)\;dt \, ,
\end{equation}
and the mean squared speed in the $i$-direction is (no sum on $i$)
\begin{equation}
\langle{\Delta v_i^2}\rangle={e^2\over m^2}\;\int_0^t\;\int_0^t\;
[\langle{E}_i({\mathbf x},t_1)\;{E}_i({\mathbf x},t_2)\rangle
-\langle{E}_i({\mathbf x},t_1)\rangle\;
    \langle{E}_i({\mathbf x},t_2)\rangle]\;dt_1\;dt_2\;.\label{eq:lang}
\end{equation}
In general, there may be a classical, nonfluctuating field in addition to the
fluctuating quantum field. However, in this case the electric field
correlation function which appears in Eq.~(\ref{eq:lang}) is just the
quantum field two-point function. Let the electric field be expressed as a sum
of a classical and a quantum part: ${\mathbf E}={\mathbf E}_c+{\mathbf E}_q$,
where $\langle {\mathbf E} \rangle = {\mathbf E}_c$. Then
\begin{equation}
\langle{\mathbf E}({\mathbf x},t_1)\;{\mathbf E}({\mathbf x},t_2)\rangle
-\langle{\mathbf E}({\mathbf x},t_1)\rangle\;
    \langle{\mathbf E}({\mathbf x},t_2)\rangle =
\langle{\mathbf E}_q({\mathbf x},t_1)\;{\mathbf E}_q({\mathbf x},t_2)\rangle\,.
\end{equation}
Thus Eq.~(\ref{eq:lang}) describes the velocity fluctuations around the
mean trajectory caused by the classical field. Henceforth, we will drop the
$q$ subscript and understand
$\langle{\mathbf E}({\mathbf x},t_1)\;{\mathbf E}({\mathbf x},t_2)\rangle$
to be the quantum two-point function.

In the presence of a boundary, this two-point function can be expressed as a
sum of the Minkowski vacuum term and a correction term due to the boundary:
\begin{equation}
\langle{\mathbf E}(x)\,{\mathbf E}(x')\rangle =
\langle{\mathbf E}(x)\,{\mathbf E}(x')\rangle_0 +
\langle{\mathbf E}(x)\,{\mathbf E}(x')\rangle_R \,,
\end{equation}
where the correction term is finite in the coincidence limit $x'=x$, so long
as the point is not actually on the boundary. The Minkowski vacuum term
would produce a formally divergent contribution to $\langle{\Delta v^2}\rangle$.
However, as discussed above, this contribution is not expected to produce any
observable consequences. Thus, we will keep only the boundary-dependent
contribution, and write
\begin{equation}
\langle{\Delta v_i^2}\rangle={e^2\over m^2}\;\int_0^t\;\int_0^t\;
\langle{E}_i({\mathbf x},t_1)\;{E}_i({\mathbf x},t_2)\rangle_R\,dt
 \; .   \label{eq:lang2}
\end{equation}

In the case of a single, perfectly reflecting plate,
$\langle{\mathbf E}(x)\,{\mathbf E}(x')\rangle_R$ can be obtained by
images~\cite{BM69}. Let the plate be located in the $z=0$ plane. At a point
a distance $z$ from the plate, the components of
$\langle{\mathbf E}({\mathbf x},t_1)\;{\mathbf E}({\mathbf x},t_2)\rangle_R$
are\footnote{Lorentz-Heaviside units with $c=\hbar =1$ will be
used here, except as otherwise noted.}
\be
 \langle{ E}_x({\mathbf x},t')\;{ E}_x({\mathbf
 x},t'')\rangle=\langle{E}_y({\mathbf x},t')\;{E}_y({\mathbf
 x},t'')\rangle=-{ \Delta t^2 + 4z^2\over \pi^2 ( \Delta t^2 -
 4z^2)^3}
\ee
and
 \be
 \langle{
 E}_z({\mathbf x},t')\;{E}_z({\mathbf
 x},t'')\rangle={ 1 \over \pi^2 ( \Delta t^2 -
 4z^2)^2}\;.
\ee

 The velocity dispersion in the $x$-direction is given by
\ben \label{eq:vx}
 \langle \Delta v_x^2\rangle=\langle \Delta v_y^2\rangle&=&{e^2\over
m^2}\;\int_0^t\;\int_0^t\;\langle{E}_x({\mathbf x},t')\;{E}_x({\mathbf
 x},t'')\rangle\; dt'\;dt''\nonumber\\
 &=& -{e^2\over \pi^2 m^2}\;\int_0^t\;\int_0^t\; {\Delta t^2 + 4z^2\over
\pi^2 ( \Delta t^2 -
 4z^2)^3}\; dt'\;dt''\nonumber\\
 &=&-{e^2\over \pi^2 m^2}\;\int_0^t\; {2(t-\tau)( \tau^2 + 4z^2)\over
\pi^2 (  \tau^2 -
 4z^2)^3}\; d\tau\;\nonumber\\
 &=&{e^2\over \pi^2 m^2}\;\left[ {t\over 64z^3}\ln\biggl( {2z+t\over
 2z-t}\biggr)^2-{ t^2\over 8z^2(t^2-4z^2)}\right]\;.
\een
  It should be pointed out that the above expression is singular at
  $t=2z$. This corresponds to a time interval equal to the
  round-trip light travel time between the particle and the plane
  boundary. Presumably, this might be a result of our assumption
  of a rigid perfectly reflecting plane boundary,  and would thus be
  smeared out in a more realistic treatment.
 For $ t\gg z$, Eq.~(\ref{eq:vx}) becomes
\be
 \langle \Delta v_x^2\rangle=\langle \Delta v_y^2\rangle\
\approx\;-{e^2\over 3\pi^2
m^2}{1\over t^2}-{8e^2\over 5 \pi^2 m^2}{z^2\over t^4}\;.
\ee
 Therefore, the mean squared velocity fluctuation in the
 directions parallel to plane decreases to zero as time approaches
 to infinity.

The velocity dispersion in the $z$-direction is
\ben
 \langle \Delta v_z^2\rangle&=&{e^2\over  m^2}\;\int_0^t\;\int_0^t\;
 \langle{E}_z({\mathbf x},t')\;{E}_z({\mathbf
 x},t'')\rangle\;dt'\;dt''\nonumber\\
 &=&{e^2\over \pi^2 m^2}\; {t\over 32z^3}\ln\biggl( {2z+t\over
 2z-t}\biggr)^2\;.
\een
 For $ t\gg z$,
\be
 \langle \Delta v_z^2\rangle\approx\;{e^2\over 4 \pi^2 m^2}{1\over z^2}
+ {e^2\over 3\pi^2 m^2}{1\over t^2}  \;. \label{eq:vz2}
\ee
Unlike the velocity dispersion in the transverse directions, that in
the direction perpendicular to the plate approaches a nonzero constant
value at late times. The fact it does not continue grow in time can be
understood as a consequence of energy conservation. Unlike the case
of Brownian motion due to thermal noise, here no dissipation is needed
for $\langle \Delta v_i^2\rangle$ to be bounded at late times.

The mean squared position fluctuation in the $x$-direction can be
calculated as follows
 \ben
 \langle\Delta x^2\rangle&=&
\int_0^t\;dt_1\;\int_0^{t_1}\;dt'\;\int_0^t\;dt_2
\;\int_0^{t_2}\;dt''\; \langle{E}_x({\mathbf x},t')\;{E}_x({\mathbf
 x},t'')\rangle\nonumber\\
&=& \frac{e^2}{\pi^2m^2}\, \left[{t^3\over192z^3}\,
\ln\left(\frac{t+2z}{t-2z}\right)^2 -{t^2\over 24z^2}-{1\over
6}\ln\left(\frac{t^2-4z^2}{4z^2}\right) \right]\;. \een For $ t\gg
z$ \be
 \langle \Delta x^2\rangle= \langle \Delta y^2\rangle\approx\;-{e^2\over
3\pi^2m^2}\ln(t/2z)\;.
\ee
The corresponding position fluctuation in the $z$ direction is
 \be
 \langle \Delta z^2\rangle= \frac{e^2}{\pi^2 m^2}\, \left[\frac{t^2}{24 z^2}
+ \frac{t^3}{96 z^3}\, \ln\left(\frac{t+2z}{t-2z}\right)^2 +
\frac{1}{6}\, \ln\left(\frac{t^2-4z^2}{4z^2}\right)\right]\; ,
 \label{eq:z2}
 \ee
and its limiting form for $ t\gg z$ is
 \be
 \langle \Delta z^2\rangle \approx \frac{e^2}{\pi^2 m^2}\, \left[
\frac{t^2}{8 z^2} + \frac{1}{3} \ln\left(\frac{t}{2z}\right) + \frac{1}{9}
+ O(z^2/t^2) \right] \;.  \label{eq:z2_large}
\ee
Recall that we have assumed that the particle does not significantly
change its position, that is, $\langle \Delta z^2\rangle \ll z^2$.
This condition will be fulfilled so long as
\be
t \ll \frac{2 \sqrt{2} \pi}{e}\, (m z)\, z \,. \label{eq:limit1}
\ee
Note that $m z$ is the ratio of the distance to the plate to the
Compton wavelength of the particle, which is typically very large.
Thus Eqs.~(\ref{eq:z2})
and (\ref{eq:z2_large}) can be valid for times long compared to $z$, the
light travel time to the plate.

Here we should also note that we have assumed no dissipation. In
the case of a ground state, such as a Casimir vacuum, it would
seem that there is no possibility of dissipation. However, we are
dealing with a situation where the interaction between the charged
particle and the quantized electromagnetic field is switched on
suddenly, and a finite time is required for the system to settle
into its steady state. During that time, dissipation of energy
suppled by the act of switching is possible. The most likely
source of dissipation here is radiation by the particle. This can
be estimated using the Larmor formula, which gives the average
power radiated by a nonrelativistic particle with acceleration $a$
to be
 \be
 P = \frac{e^2}{6 \pi}\, a^2 = \frac{e^4}{6 \pi m^2}\,
\langle {\mathbf E}^2 \rangle \, ,
\ee
where in the second step we
have used Eq.~(\ref{eq:Lang}). After a time $t$, a particle
radiating at this rate will change in its squared velocity by
 \be
\Delta v^2_{rad} = \frac{e^4 \, t}{3 \pi m^3}\, \langle {\mathbf
E}^2 \rangle = \frac{e^4 \, t}{16 \pi^3 \, z^4 \, m^3} \, ,
\ee
where we used
\be \langle {\mathbf E}^2 \rangle = \frac{3}{ 16\,
\pi^2\, z^4} \,.
\ee
The effects of radiation will be small
compared the dispersion due to vacuum fluctuations so long as
$\Delta v^2_{rad} \ll
 \langle \Delta v_z^2\rangle$, that is, so long as
\be
t \ll \frac{4\pi}{e^2}\, (m z) \, z \,.
\ee
This condition
will always be fulfilled for an electron so long as
Eq.~(\ref{eq:limit1}) holds.

\section{Interpretation of the Results}

A few comments are now in order for the above results. First, let
us notice that the Brownian motion of a test charged particle
subject to electromagnetic vacuum fluctuations will be
anisotropic, since the behaviors of both the velocity and position
dispersions are different in the longitudinal and transverse
directions. The most dramatic feature is that $\langle\Delta
v_x^2\rangle$ and $\langle \Delta x^2\rangle$ are both negative.
This is counter-intuitive and requires a physical explanation.
A negative dispersion must imply a decrease in an uncertainty
which would otherwise be present. One possibility is the usual
uncertainty in position and velocity of a quantum particle.
 Quantum mechanically, a massive
particle is described by a wave packet which must have a position
and a momentum uncertainty. It is well established that the wave
packet spreads out as time progresses, and so the position
uncertainty will increase with time. Consequently, even if the
particle is initially in a minimum uncertainty wave packet, at a
later time, it will satisfy the uncertainty principle by a wider
margin. If we recall that $\langle \Delta x^2\rangle$ is a
difference between the case with the plane boundary and that
without it, we can see that  the negative sign of $\langle\Delta
x^2\rangle$ can be understood as a reduction in the position
spreading of the wave packet in the parallel directions as
compared to what it would have been without the presence of a
plane boundary. In a somewhat different context, it has been shown
that dissipation can also suppress wave packet
spreading~\cite{SY95}. Note that the reduction due to vacuum
fluctuation is generally small as $\langle\Delta x^2\rangle$ is a
logarithmic function of time. However, the corresponding position
dispersion  in the perpendicular direction, $\langle\Delta
z^2\rangle$, is positive and furthermore it grows linearly with
time. Hence, the wave packet spreading in the $z$ direction is
reinforced by electromagnetic vacuum fluctuations and it will be
larger than what it would have been without the boundary.

Let us now discuss in more detail the wave packet spreading due to
the quantum nature of the particle and that due to
electromagnetic vacuum fluctuations. Take, as an example,  a
Gaussian wave packet which represents a particle whose position
and momentum are simultaneously determined, as closely as the
uncertainty principle permits. We will use the subscript $q$ to denote
uncertainties due to the quantum nature of a particle, as opposed to
those due to vacuum fluctuations. Assume the initial width of the
wave packet is $\Delta z_{q0}$. It can be shown that the width of the
packet at time $t$ is
\begin{equation}
\Delta z_q=\sqrt{\Delta z_{q0}^2+{(\Delta p_z)^2t^2\over m^2}}\, ,
\end{equation}
where $\Delta p_z$ is the width of the wave packet in momentum
space. Let $\Delta z_{q0} \Delta p_z= 1/2$, that is,  choose the
initial wave packet such that the uncertainty attains its
theoretical minimum value. Then
\begin{equation}
\Delta z_q=\sqrt{\Delta z_{q0}^2+{t^2\over4\Delta z_{q0}^2
m^2}}\equiv\Delta z_{qm}
\end{equation}

The question we now want to ask is: how large the position
fluctuation due to the vacuum fluctuations could be as compared to
that due to the uncertainty principle and the wave packet
spreading? For any fixed travel time $t$, we want to manipulate
the initial size of the wave packet such as at time $t$, the width
of the wave packet attains a minimum value. We find that this
initial width is
\be
\Delta z_{q0}^2={t\over 2m}\;,
 \ee
and the corresponding minimum position in any direction uncertainty is
\be
 \Delta x_{qm}= \Delta z_{qm}=\sqrt{{t\over m}}\;.
 \ee
Let
\be
\Delta x_f = \sqrt{|\langle\Delta x^2\rangle|}
\ee
be the position uncertainty in the $x$-direction due to the effects of
vacuum fluctuations, and $\Delta z_f$ be the corresponding uncertainty
in the $z$-direction. In the limit that $t \gg z$, we have, for the case
that the charged particle is an electron,
\be
\frac{\Delta x_f}{\Delta x_{qm}} =
2 \sqrt{ \frac{\alpha\, \ln(t/2z)}{3 \pi t\, m} } \, ,
\ee
where $\alpha$ is the fine structure constant. This ratio is always
very small. The corresponding ratio for the $z$-direction is
 \be
   {\Delta z_f\over \Delta z_{qm}}=\sqrt{{\alpha\over 2\pi }}\;\sqrt{t\over
m}\;{1\over
   z}=3.4\times 10^{-2}\;{\Delta z_{qm}\over
   z}\;.
   \ee
   Since the initial size, $\Delta z_{qm}$, should be
much less than
   $z$, in general this ratio is much less than one.

Note that dispersion in the transverse velocity, $\langle \Delta v_x^2\rangle$,
is essentially a transient effect which dies off rapidly in time. Although the
 dispersion in the transverse position, $\langle x^2\rangle$, grows
slowly in time, it can also be understood as consequence of the uncertainty
in $v_x$ at an earlier time, and hence also a transient effect. Such
transient effects can be due to the way in which the system is prepared.
Here we have assumed that the effect of the vacuum fluctuations begins at $t=0$
without specifying the details of how the effects are switched on. One way
this might be done is with electrons moving parallel to a finite plate and
crossing the edge of the plate at $t=0$. The effect of the switching can
cause the electron to emit photons, which can in turn contribute to
uncertainties in momentum and position. Thus, it may also be possible to
interpret negative values of $\langle \Delta x^2\rangle$ and
$\langle \Delta v_x^2\rangle$ as arising from a suppression of photon emission
effects.

The increase in $\langle \Delta v_z^2\rangle$ given in Eq.~(\ref{eq:vz2})
can be associated with an effective temperature of
 \be
   T_{eff}={\alpha\over \pi}\;{1\over k_B mz^2}=
  1.7\times 10^{-6}\left({{1 \mu m}\over z}\right)^2\;K
= 1.7\times 10^2\left({1\mathring{A}\over z}\right)^2\;K\\, , \label{eq:effT}
 \ee
where $k_B$ is Boltzmann's constant. The approximation of a perfect reflector
holds for metal plates at frequencies below the plasma frequency, which
would require that $z \agt 1 \mu m$. The corresponding temperature, although
small, is within a range that has been achieved experimentally. For
$z \alt 1 \mu m$, the plate is no longer a perfect reflector, but can be
a partial reflector. Bragg scattering can produce significant reflection
even at x-ray wavelengths.

Note that here we are concerned with the increase in the mean squared
normal velocity due to the presence of the plate. Because the electron must be
localized on a scale smaller than $z$, there is already a larger spread
in $v_z$ due to quantum uncertainty. However, in principle this could be
canceled if one measured the change due to the boundary. A free electron
near a conducting plate will also feel a classical image charge force,
but this might be canceled by another classical force. It is of interest
to compare Eq.~(\ref{eq:effT}) for an electron with the corresponding result
for an atom~\cite{WKF}, which is of order $0.1 K$ if $z = 1\mathring{A}$, and
falls as $1/z^8$ as $z$ increases. Thus the case of an electron seems
much closer to being experimentally accessible.

\begin{acknowledgments}
We would like to thank Jen-Tsung Hsiang and Ken Olum for useful discussions.
 This work was supported in part  by the National Natural Science
Foundation of China  under Grant No. 10375023, the National Basic
Research Program of China under Grant No. 2003CB71630, and by the
National Science Foundation under Grant PHY-0244898.
\end{acknowledgments}

\end{document}